\documentclass[aps, prfluids, showpacs, showkeys, notitlepage, amsmath, amssymb, floatfix, longbibliography, 12pt, tightenlines]{revtex4-2}
\usepackage[export]{adjustbox}
\usepackage{svg}
\usepackage{float}
\usepackage{amsmath}
\usepackage{amssymb}
\usepackage{multirow}
\usepackage{comment} 
\usepackage[font=small, labelfont=bf, justification=justified]{caption}
\usepackage{subfig}
\usepackage{graphicx}
\usepackage{dcolumn}
\usepackage{bm}
\usepackage{hyperref}
\DeclareMathAlphabet{\altmathcal}{OMS}{cmsy}{m}{n}
\usepackage[bottom]{footmisc}
\def\ADDA#1{{\textcolor{black}{#1}}} 
\def\ADDB#1{{\textcolor{black}{#1}}}

\begin{document}

\title{Bounds to the Basset-Boussinesq force on particle laden stratified flows}
\author{Christian Reartes\textsuperscript{*}}

\author{Pablo D. Mininni}
\affiliation{Universidad de Buenos Aires, Facultad de Ciencias Exactas y Naturales, Departamento de Física, Ciudad Universitaria, 1428 Buenos Aires, Argentina,}
\affiliation{CONICET - Universidad de Buenos Aires, Instituto de F\'{\i}sica Interdisciplinaria y Aplicada (INFINA), Ciudad Universitaria, 1428 Buenos Aires, Argentina.\\
\\
\textsuperscript{*}Author to whom correspondence should be addressed: {\rm christianreartes@df.uba.ar}}

\begin{abstract}
The Basset-Boussinesq force is often perfunctorily neglected when studying small inertial particles in turbulence. This force arises from the diffusion of vorticity from the particles and, since it depends on the particles' past history, complicates the dynamics by transforming their equations of motion into integro-differential equations. However, this force is of the same order as other viscous forces acting on the particles, and beyond convenience, the reasons for neglecting it are unclear. This study addresses the following question: Under what conditions can the Basset-Boussinesq force be neglected in light particles in geophysical flows? We derive strict bounds for the magnitude of the Basset-Boussinesq force in stably stratified flows, in contexts of interest for geophysical turbulence. The bounds are validated by direct numerical simulations. The Basset-Boussinesq force is negligible when a buoyancy Stokes number $\textrm{Sb} = N \tau_p$ is small, where $N$ is the flow Brunt-V\"ais\"al\"a frequency and $\tau_p$ is the particle's Stokes time. Interestingly, for most oceanic particles this force may be negligible. Only for very strong stratification, or for particles with very large inertia, this force must be considered in the dynamics.
\end{abstract}

\maketitle

\section{Introduction}

The dynamics of inertial particles submerged in turbulent flows plays a crucial role in various geophysical contexts, from phenomena in coastal environments and lakes, to the atmosphere and the oceans \cite{wyngaard_1992, dasaro_2000, watanabe_2017, amir_2017, Ichihara_2023}. Despite extensive research on the influence of turbulence on the transport and spatial distribution of particles, a complete description of their dynamics in geophysical systems remains elusive. 
Furthermore, the complexity increases in the environmental sphere, where the dynamics of plankton and algae, and the presence of microplastics in the oceans, introduce additional layers of difficulty to particle dispersion, and emphasize the urgent need to understand these phenomena \cite{OBRIEN_2023}. In many of these cases the complexity of modeling and simulating these systems imposes the need to use reduced models that simplify the physics. Recently, both experiments \cite{Obligado_2015, sofi_2022} and particle-resolved simulations \cite{Tavanashad_2021, Vowinckel_2023} have provided valuable insights into particle dynamics in different regimes. Many hydrodynamic forces operate over the particles at small scales within turbulent flows, influencing their interactions, collisions, and cluster formation \cite{Tavanashad_2021}. 

The modeling of geophysical flows, particularly those featuring stable density stratification, presents also distinct challenges even in the absence of particles. Anisotropy characterizes stably stratified turbulence, setting it apart from homogeneous isotropic turbulence (HIT) \cite{Riley_2003, lindborg_2008, marino_2014, Portwood_2019, Alexakis_2024}. Within these flows, the impact of stratification is evident in the reduction of vertical velocity, leading to confined, quasi-horizontal layered motions, and to the generation of vertically sheared horizontal winds marked by significant vertical variability \cite{smith_2002}. The stratified environment introduces a restoring force, allowing for the coexistence of waves with turbulence, each exhibiting a distinct spectral scaling compared to HIT. Stably stratified turbulence manifests an anisotropic inertial subrange, fostering a direct energy cascade between buoyancy and Ozmidov scales \cite{waite_2011, maffioli_2017}. Notably, studies also suggest that larger-scale quasi-horizontal motions serve as a continuous source of small-scale turbulence, provided the local Reynolds number does not fall below a critical threshold \cite{Riley_2003}. Examining these phenomena, it becomes apparent that in stably stratified turbulence, as described by Herring and M\'etais \cite{Herring_1989} and by Riley and Lelong \cite{riley_lelong}, thin layers of large quasihorizontal vortical structures coexist with internal gravity waves \cite{brethouwer_2007}. All these features underscore the complex interplay of forces and structures characterizing turbulence in geophysical flows.

In this context, several recent studies have considered the dynamics of different particles advected by stable stratified flows. As a first example, analysis of Lagrangian tracers in stratified turbulence was performed in \cite{van_aartrijk_2008, Sujo_2018}. When particles have inertia, for small-sized particles in a turbulent flow the Maxey-Riley-Gatignol approximation provides a set of equations to describe their dynamics \cite{1,Gatignol_1983}. This approximation was \ADDA{either used heuristically, or} extended to stratified flows, for heavy \cite{van_aartrijk_2010} and for neutrally buoyant and light particles \cite{van_aartrijk_2010, Sozza2016, reartes_2023} \ADDA{(see \cite{reartes_2023} for a derivation of equations for inertial particles compatible with the Boussinesq approximation in stratified fluids)}. As particles increase in size, Fax\'en corrections proportional to the square radius of the particles become relevant \cite{van_aartrijk_2010}. But even for very small particles the Maxey-Riley-Gatignol equation is an integro-differential system for the particles' evolution that depends on the particles' past history. To further simplify the problem the Basset-Boussinesq force, resulting from the diffusion of vorticity away from the particles along their trajectories, is often neglected. Methods to reduce the computational cost to estimate this force were devised \cite{VANHINSBERG20111465}, and a few studies have considered its effect in particle laden stratified flows \cite{van_aartrijk_2010}. It was seen that this force becomes particularly relevant in the presence of flow stratification, as particles experience significant acceleration and deceleration resulting from oscillatory motions caused by buoyancy. However, it is still unclear under what general conditions this force becomes dominant, or negligible.

\ADDB{As an example of an application of the study of particle-laden flows in geophysical contexts, we can mention models based on heuristically modified Maxey-Riley-Gatignol equations, such as the one used by Beron-Vera et al.~for the transport of Sargassum, marine debris, and garbage patches in the ocean \cite{beron_vera_sep_2020, beron_vera_nov_2020, beron_vera_2021}. The authors modify the Maxey-Riley-Gatignol equation to account for the effect of the Coriolis force and oceanic vorticity, leading to a specific formulation for the evolution of inertial particles that incorporates these effects. Another example is provided by the work in \cite{Sozza_2018}, which empirically models the dynamics of floaters, such as plankton in the ocean, by considering fluid stratification while neglecting rotation. Both models highlight the importance of considering inertial effects and drag forces, as well as environmental forces, in the accurate description of particle dynamics in geophysical flows \cite{beron_vera_jul_2020}. But both models also neglect the Basset-Boussinesq history force.}

In this work we derive a bound to the Basset-Boussinesq force in a stratified fluid, that allows estimation of conditions under which this force can be neglected. To obtain this bound we consider the Maxey-Riley-Gatignol equation for small inertial particles in a stably stratified fluid (see Sec.~\ref{sec:theory}, where we also discuss its range of applicability and typical values of parameters in geophysical flows, and Sec.~\ref{sec:bound} where we present the bound). We also conduct direct numerical simulations of the Navier-Stokes equation under the Boussinesq approximation for the fluid, along with the Maxey-Riley-Gatignol equation with and without the Basset-Boussinesq (or history) term (see Sec.~\ref{sec:setup}). In Sec.~\ref{sec:results} we compare particle dispersion and particle velocity statistics, as well as the particle preferential clustering, for different values of the controlling parameters. We show that the bound allows estimation of under what conditions neglecting the Basset-Boussinesq history term does not affect the statistical behavior of any of these observables (see Sec.~\ref{sec:conclusions} for the main conclusions).

\section{Equations of motion \label{sec:theory}}

We consider a stably stratified fluid described by the incompressible Navier-Stokes equation under the Boussinesq approximation for the velocity ${\bf u} = (u,v,w)$, and an equation for mass density fluctuations $\rho'$,
\begin{equation}
\partial_t {\bf u} + {\bf u} \cdot \boldsymbol{\nabla} {\bf u} = - \boldsymbol {\nabla} \left(p/\rho_0\right) - \left(g/\rho_0 \right)\rho' \hat{z}+ \nu \nabla^2 {\bf u} + {\bf f},
\end{equation}
\begin{equation}
\partial_t \rho'+ {\bf u} \cdot \boldsymbol{\nabla} \rho' = \left(\rho_0 N^2/g\right) w + \kappa \nabla^2 {\bf \rho'},
\end{equation}
\begin{equation}
\nabla \cdot {\bf u} = 0,
\end{equation}
where $p$ is the correction to the hydrostatic pressure, $\nu$ is the kinematic viscosity, ${\bf f}$ is an external mechanical forcing, $N$ is the aforementioned Brunt-V\"{a}is\"{a}l\"{a} frequency (which in this approximation sets the stratification), and $\kappa$ is the diffusivity. In terms of the background density gradient, the Brunt-V\"{a}is\"{a}l\"{a} frequency is $N^2 = -(g/\rho_0) (d\bar \rho /dz)$, with $d \bar \rho /dz$ the imposed linear background stratification, and $\rho_0$ the mean fluid density. We write scaled density fluctuations $\zeta$ in units of velocity by defining $\zeta = g\rho'/(\rho_0 N)$.  All quantities are then made dimensionless using a characteristic length $L_{0}$ and a characteristic velocity $U_{0}$ in the domain, resulting in
\begin{equation}
\partial_t {\bf u} + {\bf u} \cdot \boldsymbol{\nabla} {\bf u} = - \boldsymbol {\nabla} \left(p/\rho_0\right) - N\zeta \hat{z}+ \nu \nabla^2 {\bf u} + {\bf f},
\label{bou1}
\end{equation}
\begin{equation}
\partial_t \zeta + {\bf u} \cdot \boldsymbol{\nabla} \zeta = N w + \kappa \nabla^2 {\bf \zeta}.
\label{bou2}
\end{equation}

\ADDB{We describe small neutrally buoyant inertial particles using the Maxey-Riley-Gatignol equation, under an approximation consistent with those made in the Boussinesq approximation used to describe the fluid (see \cite{reartes_2023} for a derivation assuming that particles are sufficiently small and thus that the flow in the vicinity of the particles is homogeneous). Note that} under the Boussinesq approximation for a stratified fluid, Eqs.~(\ref{bou1}) and (\ref{bou2}) are obtained from the Navier-Stokes equations after neglecting density fluctuations everywhere except in the buoyancy (or gravity) force. Thus, for the dynamics of the particles we also approximate the density and the mass of the fluid displaced by the particles by their mean values, i.e., $\rho_f  \approx \bar{\rho}_f = \rho_0$ and $m_f \approx \bar{m}_f = \rho_0 V_p$ (where $V_p$ is particle volume), except in the gravity term. In that term we keep the complete fluid density dependence, $\rho_f = \rho_0 + (d\bar{\rho}/d z)(z-z_0)+\rho'$ (note that for a stably stratified fluid $d\bar{\rho}/d z < 0$). Therefore, the full effect of gravity and buoyancy forces acting on the particles is preserved. For simplicity we also assume that the typical length over which the velocity field changes appreciably is much larger than the particle radius $a$, and Fax\'en corrections are thus neglected. With these approximations the equation for the \ADDB{neutrally buoyant particles} is
\begin{equation}
\Dot{\bf v} = \frac{\left[ {\bf u}({\bf x}_p,t) - {\bf v}(t) \right]}{\tau_p} - \frac{2N}{3} \left[N(z_p-z_0) - \zeta \right] \hat{z} + \frac{\textrm{D}}{\textrm{D}t} {\bf u}({\bf x}_p,t) +\sqrt{\frac{3}{\pi \tau_p}}\int_{-\infty}^{t} d\tau \frac{\frac{\textrm{d}}{\textrm{d}\tau} [{\bf u}({\bf x}_p,\tau) - {\bf v}(\tau)]}{\sqrt{t-\tau}} ,
\label{ec.parts}
\end{equation}
where ${\bf x}_p(t)$ is the particle position, ${\bf v}(t)$ is the particle velocity, ${\bf u}({\bf x}_p,t)$ is the fluid velocity at the particle position, $D/Dt$ is the Lagrangian derivative, $\textrm{d}/\textrm{d}t$ is the time derivative following the particle trajectory, $z_0$ is the height at which particles are neutrally buoyant, \ADDB{$\zeta$ is the fluid density fluctuation at the particle position}, and  $\tau_p = (m_p + \bar{m}_f/2)/(6 \pi a \bar{\rho}_f \nu)$ is the particle relaxation time. For a spherical particle $\tau_p = a^2/(3 \nu)$ when $\bar{m}_f/m_p = 1$ (note that this choice does not make the dynamics independent of the ratio of densities of the fluid and the particles; under these approximations any value of $\bar{m}_f/m_p \neq 1$ is equivalent to changing the reference value $\rho_0$ of the fluid, and results in particles being neutrally buoyant at a different height, which can in turn be absorbed into a new reference value of $z_0$).

The second term on the r.h.s.~of Eq.~(\ref{ec.parts}) 
corresponds to gravity and buoyancy forces. Indeed, when multiplied by $m_p$, $-(2/3) \, m_p N^2 \, (z_p - z_0) = (2/3) \, m_p g \, (\rho_0^{-1} d\bar{\rho}/dz) (z_p - z_0)$, and $(2/3) \, m_p N \zeta = (2/3) \, m_p g  \, \rho'/\rho_0$. Vertical forces acting on the particle thus depend on the height at which the particle is (the fluid background stratification, which is negative, pushes heavier particles downwards and lighter particles upwards), as well as on local density fluctuations of the fluid density (variations in the total fluid density associated with $\rho'$ affect the buoyancy forces felt by the particles). Gravity and buoyancy can thus be the most important forces in the vertical dynamics of the particles. Note a similar role is played in the horizontal particles' velocities in \ADDA{the heuristic model} in \cite{Beron_2019, BeronVera2020} by the Coriolis force. 

We finally define the particles Stokes number as $\textrm{St} = \tau_p/\tau_\eta$, where $\tau_\eta = (\nu/\varepsilon)^{1/2}$ is the Kolmogorov time scale, and $\varepsilon$ is the fluid kinetic energy dissipation rate.
With this definition we can discuss typical values of this number and of dimensional parameters of particles in geophysical contexts, which will be useful in Secs.~\ref{sec:bound} and \ref{sec:setup}. A typical value of the Brunt-V\"ais\"al\"a frequency in oceanic flows is $N \approx 10^{-3}$ s$^{-1}$. The oceanic lower mixed layer or the upper seasonal thermocline have a mean Kolmogorov time scale $\tau_\eta \approx 3$ s, and Kolmogorov dissipation scales between $0.1$ and 10 mm \cite{Durham_2013}. Nanoplankton and microplankton have typical sizes between $2$ to $200$ microns \cite{AcevedoTrejos2015}, with Stokes numbers varying between $\textrm{St} \approx 10^{-5}$ to $10^{-1}$. Other neutrally buoyant particles in the ocean have particle radii between $1$ to $50$ mm, and particle response times between $\tau_p\approx 0.1$ to $100$ s \cite{Squires_1995}, resulting in Stokes numbers between $\textrm{St} \approx 0.3$ to $30$. The mass ratio for oceanic neutrally buoyant particles can be as close to unity as $\approx 1.003$ \cite{McCave_1984}. Sargassum, seaweed, and marine debris in upper oceanic layers can have larger sizes, but are still modeled with modified Maxey-Riley-Gatignol equations using effective Stokes numbers smaller than 1 \cite{Beron_2019, BeronVera2020}. Even though it is not the focus of this study, we also briefly mention that in the atmosphere the Brunt-V\"ais\"al\"a frequency varies between $0.1$ to $0.01$ s$^{-1}$ depending on altitude and atmospheric conditions \cite{nozawa_2023}, while light particles can have response times on the order of $0.05$ s \cite{Shaw_2003}.

\section{Bounds to the Basset-Boussinesq force \label{sec:bound}}

The integral in the Basset-Boussinesq history term in Eq.~(\ref{ec.parts}) can be written as a convolution between a function $\textbf{g}(t)$ and a kernel $K_{\textrm{BB}}(t)$, $\int_{-\infty}^t K_{\textrm{BB}}(t-\tau) \textbf{g}(\tau) d \tau$, where
\begin{eqnarray}
\textbf{g}(t) = \frac{\textrm{d}\textbf{h}(t)}{\textrm{d}t}~,~~~~~~ \textbf{h}(t)={\bf u}({\bf x}_p(t),t) - {\bf v}(t)={\bf v}_\textrm{slip}(t)~,~~~~~~
K_{\textrm{BB}}(t)=\frac{1}{\sqrt{t}}~, \label{em3}
\end{eqnarray}
Note that $\textbf{h}(t)$ is the slip velocity of each particle only as long as Faxén corrections are neglected. Both the Stokes term in Eq.~(\ref{ec.parts}), $\textbf{h}(t)/\tau_p$, and the Basset-Boussinesq history term, depend on the particle inertia. Thus, it is reasonable to bound the Basset-Boussinesq force compared against the Stokes force. 

In the most general (not stratified) case, Van Hinsberg et al.~\cite{VANHINSBERG20111465} proved that the Basset-Boussinesq force is finite at any given time, provided some constraints on $\textbf{h}(t)$ and $\textbf{g}(t)$ are satisfied. First, $\textbf{h}(t)$ must be a continuous function, and its derivative must exist almost everywhere. Second, the infinity norm of $\textbf{h}(t)$ and $\textbf{g}(t)$ must be finite. These constraints on $\textbf{h}(t)$ and $\textbf{g}(t)$ are
\begin{eqnarray}
\textbf{h}\in C^0, ~~~~~\|\textbf{h}\|_{\infty}=B_1,
~~~~~\|\textbf{g}\|_{\infty}=B_2,\label{cond1}
\end{eqnarray}
where $\|\cdot\|_{\infty}$ is defined as
\begin{eqnarray}
\|\textbf{h}\|_{\infty}= \inf \{ C\ge 0:|\textbf{h}(t)| \le C \mbox{
almost anywhere}\},\label{cond2}
\end{eqnarray}
where $|\cdot|$ represents the typical vector length. Let's assume that these conditions are fulfilled for particles in turbulent flows with $\textbf{h}(t)={\bf v}_\textrm{slip}(t)$. Under these conditions an upper bound for the Basset-Boussinesq force $\textbf{F}_\textrm{BB}$ can be obtained. The convolution is separated into two parts to control both the singularity in the kernel and in the tail of the integral \cite{VANHINSBERG20111465}:
\begin{eqnarray}
\frac{|\textbf{F}_\textrm{BB}|}{c_\textrm{BB}}&=&
\left| \int_{-\infty}^tK_{\textrm{BB}}(t-\tau)
\textbf{g}(\tau)\textrm{d}\tau
\right|\nonumber\\
&=&\left|
\int_{-\infty}^{t-B_1/B_2}\frac{\textbf{g}(\tau)}{\sqrt{t-\tau}}
\textrm{d}\tau+\int_{t-B_1/B_2}^t\frac{\textbf{g}(\tau)}{\sqrt{t-\tau}}\textrm{d}\tau
\right|\nonumber\\
&\leq&\left|\left[ \frac{
\textbf{f}(\tau)}{\sqrt{t-\tau}}\right]_{-\infty}^{t-B_1/B_2}-\int_{-\infty}^{t-B_1/B_2}\frac{
\textbf{f}(\tau)}{2(t-\tau)^{3/2}}\textrm{d}\tau\right|+\int_{t-B_1/B_2}^t\frac{\left|\textbf{g}(\tau)\right|}{\sqrt{t-\tau}}\textrm{d}\tau
\nonumber\\
&\leq&\sqrt{B_1B_2}+\frac{B_1}{2}\int_{-\infty}^{t-B_1/B_2}\frac{
1}{(t-\tau)^{3/2}}\textrm{d}\tau+B_2\int_{t-B_1/B_2}^t\frac{1}{\sqrt{t-\tau}}\textrm{d}\tau
\nonumber\\
&=& 4\sqrt{B_1B_2},
\label{cot_van_hin}
\end{eqnarray}
where $c_\textrm{BB}=\sqrt{3/(\pi \tau_p)}$, and where this bound holds for all times (note that while this separation can also be used to implement a numerical scheme for the Basset-Boussinesq force, as in \cite{van_Hinsberg_2011}, this result uses the separation only to bound the force and does not depend on the method used to estimate the force numerically).

In the stratified case, Eq.~(\ref{ec.parts}) can be rewritten as
\begin{equation}
\ddot{\bf x}_p + \frac{1}{\tau_p} \dot{\bf x}_p + \frac{2}{3} N^2 z_p \hat{z} = {\bf F}(t),
\label{eq:osc}
\end{equation}
\ADDB{where
\begin{equation}
{\bf F}(t) = \frac{{\bf u}({\bf x}_p,t) }{\tau_p} + \frac{2N}{3} \left[N z_0 + \zeta \right] \hat{z} + \frac{\textrm{D}}{\textrm{D}t} {\bf u}({\bf x}_p,t) +\sqrt{\frac{3}{\pi \tau_p}}\int_{-\infty}^{t} d\tau \frac{\frac{\textrm{d}}{\textrm{d}\tau} [{\bf u}({\bf x}_p,\tau) - \dot{\bf x}_p(\tau)]}{\sqrt{t-\tau}} ,
\label{ext_force}
\end{equation}
is a force acting on each particle. This force ${\bf F}$ should not be confused with the external force ${\bf f}$ acting on the fluid in Eq.~(\ref{bou1}); note that Eq.~(\ref{ext_force}) includes all the forces exerted on the particle by the fluid in Eq.~(\ref{ec.parts}).}

In the vertical component, Eq.~(\ref{eq:osc}) is an equation for a driven damped oscillator with frequency $\sqrt{2/3}N$ and with damping constant $1/(2\tau_p)$ \cite{reartes_2023}. As in the vertical direction we have fast fluctuations of particles velocities, it makes sense to continue looking at this component of the equation. In \cite{reartes_2023} it was shown that, assuming that in a stratified flow the vertical fluid velocity is dominated by internal gravity waves, and thus approximating fluid element displacements by those resulting from the propagation of these waves, $z_f = z_0 + \zeta/N$, $\dot{z}_f=w({\bf x}_p(t),t)$, and $\ddot{z}_f = Dw/Dt$, we can write the vertical component of Eq.~(\ref{ec.parts}) as
\begin{equation}
(\ddot{z}_p - \ddot{z}_f) + \frac{1}{\tau_p} (\dot{z}_p - \dot{z}_f) + \frac{2}{3} N^2 (z_p - z_f)  = {\bf F}_\textrm{BB} \cdot \hat{z}.
\end{equation}
\ADDB{The vertical component of the first three terms in the force ${\bf F}$ are now absorbed on the l.h.s.~of this equation, leaving only ${\bf F}_\textrm{BB}$ on the right.} The homogeneous solution to this equation, neglecting damping (as we are looking for maximum bounds for the velocities and viscous forces acting on the particle), is $z_p - z_f = A_0 \exp(i\sqrt{2/3}Nt)$, and thus we can approximate the vertical slip velocity as $w_\textrm{slip} = W_0 \exp(i\sqrt{2/3}Nt)$ (real parts are assumed everywhere). The vertical component of the Stokes force can then be written as,
\begin{equation}
{\bf F}_\textrm{St} \cdot \hat{z} = \frac{W_0}{\tau_p} e^{i\sqrt{2/3} Nt} .
\label{fst}
\end{equation}
Replacing the vertical slip velocity in the vertical component of the Basset-Boussinesq force we also obtain
\begin{equation}
{\bf F}_\textrm{BB} \cdot \hat{z} = \sqrt{\frac{3}{\pi \tau_p}}\int_{-\infty}^{t} d\tau K_{\textrm{BB}}(t-\tau) \, i\sqrt{\frac{2}{3}} N W_0 e^{i\sqrt{2/3}N\tau},
\label{fbb}
\end{equation}
where we used that the vertical component of the convolution function in the history term is $g_z(t) = \textrm{d} w_\textrm{slip}/\textrm{d}t = i\sqrt{2/3} N w_\textrm{slip}(t)$.

From these results, we can proceed further in deriving bounds for the different forces over particles in the stratified case using Eqs.~(\ref{em3}), (\ref{cond2}), (\ref{fst}), and (\ref{fbb}). As vertical velocity fluctuations are faster and their time derivatives are larger than in the horizontal direction, we can keep working with the vertical components of the Stokes and Basset-Boussinesq forces. We can ask the conditions on Eq.~(\ref{cond2}) to apply to the vertical component, and thus
\begin{equation}
\|{\bf F}_\textrm{St} \cdot \hat{z}\|_{\infty}=\left\Vert \frac{W_0}{\tau_p}  e^{i\sqrt{2/3} Nt} \right\Vert_{\infty} = \frac{B_{1z}}{\tau_p},
\label{relation_1}
\end{equation}
\begin{equation}
B_{2z} = \left\Vert i \sqrt{\frac{2}{3}} N W_0 e^{i\sqrt{2/3} Nt} \right\Vert_{\infty} = \sqrt{\frac{2}{3}} N B_{1z}.
\label{relation_2}
\end{equation}
Applying the bound on Eq.~(\ref{cot_van_hin}) only to the vertical component of the Basset-Boussinesq force, we can write a new bound as
\begin{eqnarray}
\|{\bf F}_\textrm{BB} \cdot \hat{z}\|_{\infty} 
&\leq& 4c_\textrm{BB} \sqrt{B_{1z}B_{2z}}
= 4\sqrt{\sqrt{\frac{2}{3}} \, \frac{3N}{\pi \tau_p} \left(\tau_p \|{\bf F}_\textrm{St} \cdot \hat{z}\|_{\infty}\right)^2}\leq 4 \sqrt{N \tau_p} \|{\bf F}_\textrm{St} \cdot \hat{z}\|_{\infty},
\end{eqnarray}
and therefore
\begin{eqnarray}
\frac{\|{\bf F}_\textrm{BB} \cdot \hat{z}\|^{2}_{\infty}}{\|{\bf F}_\textrm{St} \cdot \hat{z}\|^{2}_{\infty}}
&\leq& 16 N\tau_p.
\label{cota}
\end{eqnarray}
Assuming that for sufficiently stratified flows the vertical components of the forces acting on the particles dominate their dynamics, we can define a dimensionless number
\begin{equation}
\textrm{Sb} = N\tau_p ,
\label{Sb}
\end{equation}
which in the following we will call the buoyancy Stokes number, as it corresponds to a ratio between the particle response time and the Brunt-V\"ais\"al\"a period \footnote{It is possible to show that $\textrm{Sb} \sim \textrm{St}/\textrm{Rb}^{\frac{1}{2}}$, which provides a measure of the particles' inertia relative to the intensity of turbulence at the buoyancy scale.}. We expect that for sufficiently small $\textrm{Sb}$ the effect of the Basset-Boussinesq force should be negligible.

It is important to remark that the bound given by Eq.~(\ref{cota}) can only be valid when the stratification of the fluid is sufficiently strong, and cannot hold in the limit of an isotropic and homogeneous fluid with $N \to 0$. This is a direct consequence of Eq.~(\ref{eq:osc}) being that of a driven damped oscillator only when the fluid is stratified. Later, when deriving Eq.~(\ref{relation_1}), we also assumed vertical fluid velocity fluctuations are faster than horizontal velocity fluctuations, which amounts to assuming that the fluid Froude number $\textrm{Fr} = 1/(T N) < 1$, where $T$ is some characteristic fluid turnover time (see Sec.~\ref{sec:setup} for detailed definitions).

\section{Numerical simulations \label{sec:setup}}

We now focus our attention on the numerical validation of the condition obtained in Sec.~\ref{sec:bound} for regimes in which the Basset-Boussinesq force can be neglected or not. To this end we performed several numerical simulations of stably stratified turbulence (see table \ref{tab_fluid}), each of them with different particles (see table \ref{tab_parts}). The Boussinesq fluid equations given by Eqs.~(\ref{bou1}) and (\ref{bou2}) were numerically solved in a triply periodic domain using a parallelized and fully dealiased pseudo-spectral method, along with a second-order Runge-Kutta scheme for time integration \cite{mininni_hybrid_2011}. For the evolution of inertial particles satisfying Eq.~(\ref{ec.parts}) we used third-order spline interpolation to estimate forces at the particles positions, and a second-order Runge-Kutta method for time integration \cite{Yeung_1988}. The Basset-Boussinesq force was computed using the second-order method described in van Hinsberg et al.~\cite{VANHINSBERG20111465}. We explicitly verified that errors in this numerical integration scheme for the Basset-Boussinesq force remained small, by comparing against explicit integration of a small fraction of the particles in simulations with the same fluid and particle parameters.

\begin{table*}
\caption{\label{tablaga} Relevant parameters of the fluid simulations. $NT_0$ is the Brunt-Väisälä frequency in units of $T_0^{-1}=U_0/L_0$, $\textrm{Fr}$ is the Froude number, $\textrm{Re}$ is the Reynolds number, $\textrm{Rb}$ is the buoyancy Reynolds number, $L$ is the flow integral scale, $\eta$ is the Kolmogorov scale, $L_b$ is the buoyancy length, and $L_{Oz}$ is the Ozmidov length scale. All lengths are in units of the unit length $L_0$.}
\begin{ruledtabular}
\begin{tabular}{ccccccccc}
 Run  & $N T_0$ & $\textrm{Fr}$ & $\textrm{Re}$ & $\textrm{Rb}$ & $L/L_0$ & $\eta/L_0$ & $L_b/L_0$ & $L_{Oz}/L_0$ \\ \hline
N04  & 4  & 0.19 & 3600 & 130 & 1.22 & 0.0045 & \ADDB{1.51} & 0.35 \\ 
N08  & 8  & 0.11 & 2300 &  28 & 0.90 & 0.0050 & \ADDB{0.63} & 0.17 \\
N12  & 12 & 0.07 & 2500 &  12 & 0.97 & 0.0051 & \ADDB{0.44} & 0.12 \\
N20  & 20 & 0.05 & 2400 &   6 & 0.91 & 0.0048 & \ADDB{0.25} & 0.07
\label{tab_fluid}
\end{tabular}
\end{ruledtabular}
\end{table*}

\begin{table*}
\caption{\label{tablaga} Parameters of the particles in all simulations. $\textrm{St}$ is the Stokes number, $\tau_p/T_0$ is the Stokes time in units of $T_0$, $a/\eta$ is the particle radius in units of the Kolmogorov scale, and $\textrm{Re}_p$ lists the particle Reynolds numbers and $\textrm{Sb}$ the buoyancy Stokes number in all fluid simulations.}
\begin{ruledtabular}
\begin{tabular}{cccccccccccc}
\multirow{2}{*}{Label} & \multirow{2}{*}{$\textrm{St}$} & \multirow{2}{*}{$\tau_p/T_0$} & \multirow{2}{*}{$a_p/\eta$} & \multicolumn{4}{c}{$\textrm{Re}_p$} & \multicolumn{4}{c}{$\textrm{Sb}$}\\
\cline{5-8}\cline{9-12}
   &   &   &   & N04 & N08 & N12 & N20 & N04 & N08 & N12 & N20 \\
\hline
St03  & 0.3 & 0.02 & 0.95 & 0.19 & 0.15 & 0.07 & 0.05 & 0.08 & 0.19 & 0.28 & 0.43\\ 
St1   & 1   & 0.07 & 1.70 & 0.72 & 0.52 & 0.22 & 0.12 & 0.25 & 0.64 & 0.94 & 1.40\\ 
St3   & 3   &   0.21 & 3.00 & 2.70 & 2.00 & 0.75 & 0.40 & 0.75 & 1.90 & 2.80 & 4.30\\
St6   & 6   &   0.43 & 4.20 & 6.70 & 4.80 & 1.70 & 0.33 & 1.50 & 3.80 & 5.60 & 8.60
\label{tab_parts}
\end{tabular}
\end{ruledtabular}
\end{table*}

Numerical simulations were carried out using a spatial resolution of $N_x = N_y = 768$ and $N_z = 192$ grid points. The domain had dimensions $L_x=L_y=2\pi L_0$ in the horizontal directions, and $L_z=H=\pi L_0/2$ in the vertical direction, where $L_0$ is a unit length. We considered four different Brunt-Väisälä frequencies (see table \ref{tab_fluid}), measured in units of the inverse of a unit turnover time $T_0 = L_0/U_0$, with $U_0$ representing a unit velocity. For simplicity, all simulations had a Prandtl number $\textrm{Pr} = \nu/\kappa = 1$. The kinematic viscosity was chosen such that the Kolmogorov scale $\eta = (\nu^3/\varepsilon)^{1/4} \approx 0.005 L_0$ was well resolved, where the kinetic energy dissipation rate is $\varepsilon = \nu \left< |\boldsymbol{\omega}|^2\right>$ and $\boldsymbol{\omega} = \boldsymbol{\nabla} \times {\bf u}$ is the vorticity. This results in \ADDB{$\kappa_\textrm{max} \eta \approx 1.6$, where $\kappa_\textrm{max} = N_x/(3L_0)$} is the maximum resolved wave number when using the $2/3$ rule for dealiasing, ensuring spatially well-resolved simulations \cite{Donzis_2010, Wan_2010}.

The forcing in Eq.~(\ref{bou1}) was a Taylor-Green forcing, that excites directly the flow horizontal velocity components, and produces large-scale counter-rotating vortices perpendicular to the stratification separated by horizontal shear layers in between. Its expression is given by
\begin{equation}
    {\bf f} = f_{0} \left[ \sin(k_{f}x)\cos(k_{f}y)\cos(k_{f}z) \hat{\bf x} - \cos(k_{f}x)\sin(k_{f}y)\cos(k_{f}z) \hat{\bf y} \right],
\end{equation}
where $f_0$ is the forcing amplitude and $k_f=1/L_0$ is the forcing wave number. This forcing has been used in many studies of stratified turbulence (see, e.g., \cite{Riley_2003, Sujo_2018} for detailed discussions of the flow geometry and for visualizations). Note that this forcing remains constant over time, introducing no additional time scales into the system.

Particles were initialized randomly in a horizontal strip of width $H/5$, centered around $z_0 = H/2$, and at a time at which the flows had reached a turbulent steady state (i.e., after flow integration for over 60 large-scale turnover times). The initial velocities of the particles matched the fluid velocity at the center of each particle. The particles were one-way coupled, essentially functioning as test particles. They neither collide with each other nor back react into the flow dynamics. In each fluid simulation in table \ref{tab_fluid} four sets of particles were introduced (see table \ref{tab_parts}), each containing 25,000 particles, each set characterized by distinct values of $\tau_p$. Particles were integrated for more than 15 large-scale turnover times \ADDB{(simulations with particles with $\textrm{St} = 6$ were extended for twice this time, as particles with more inertia have a longer relaxation time; see below for a definition)}. Moreover, for each set of particles, numerical integrations were done solving Eq.~(\ref{ec.parts}) with and without the Basset-Boussinesq history term (labed in the following respectively as ``w/H" and ``w/oH"). This resulted in a cumulative count of 32 particle datasets, each with their corresponding Reynolds, Froude, and Stokes numbers as defined next.

We can characterize the flow dynamics using two dimensionless numbers, the Reynolds and Froude numbers,
\begin{equation}
 \textrm{Re} = \frac{LU}{\nu}, \,\,\,\,\,\, 
 \textrm{Fr} = \frac{U}{LN},
 \label{eq:Re_Fr}
\end{equation}
where $L=\pi/(2u'^2) \int E(k)/k \, dk$ and $U=\langle |{\bf u}|^2 \rangle^{1/2}$ are respectively the characteristic Eulerian integral length and the r.m.s.~flow velocity (where $E(k)$ is the isotropic kinetic energy spectrum, and $u'^2=U^2/3$). A turnover time can then be defined as $T=L/U$. With the aid of these dimensionless numbers, we can also define the buoyancy Reynolds number,
\begin{equation}
 \textrm{Rb}  =   \textrm{Re} \, \textrm{Fr}^{2},
\end{equation}
which quantifies the turbulence intensity at the \ADDB{buoyancy scale $L_{b}=2\pi/k_{b}$, with $k_{b} = N/U$,} and plays a crucial role in describing the flow behavior. When $\textrm{Rb} \gg 1$ turbulence is strong even in the presence of stratification, while for $\textrm{Rb} \ll 1$ turbulent motions are significantly suppressed by viscosity in each stratified layer. Geophysical flows typically exhibit large values of $\textrm{Rb}$ \cite{moum_1996}, but computational constraints impose severe restrictions on the range of values of $\textrm{Rb}$ that can be simulated. Previous studies indicate that $\textrm{Rb}>10$ is sufficient for the flow to sustain significant turbulence at small scales \cite{Ivey2008}. Considering computational limitations and the need to explore parameter space, we consider flows with $\textrm{Rb}$ ranging from 6 to 130. Another relevant length scale to characterize the small scales of stratified turbulence is the Ozmidov scale, $L_{Oz}=2\pi/k_{Oz}$, with $k_{Oz}=(N^3/\varepsilon)^{1/2}$. At scales significantly smaller than $L_{Oz}$ we expect the flow to asymptotically recover isotropy. Thus, when $\textrm{Rb}$ is sufficiently large and $L_{Oz}$ is larger than the the Kolmogorov dissipation scale, we can expect small-scale turbulence to be stronger, and to affect the particle dynamics and transport.

The particles dynamics is usually characterized in turn by two dimensionless numbers,
\begin{equation}
 \textrm{St} = \frac{\tau_p}{\tau_\eta}, \,\,\,\,\,\, 
 \textrm{Re}_p = \frac{a|\bf{u}-\bf{v}|}{\nu},
\end{equation}
where $\textrm{St}$ is the aforementioned Stokes number, $\tau_\eta$ is the Kolmogorov dissipation time defined before, and $\textrm{Re}_p$ is the particle Reynolds number.

Tables \ref{tab_fluid} and \ref{tab_parts} provide all these dimensionless numbers and characteristic scales for the simulations. Note these tables should be read together, as we can have, e.g., particles with $\textrm{St}=0.3$ in a flow with $N=4/T_0$ (with or without the Basset-Boussinesq history term), or the same particles but in a flow with $N=8/T_0$, $12/T_0$, or $20/T_0$. To contextualize the choice of parameters, note that in oceanic flows the Froude number varies from $\textrm{Fr} \approx 10^{-2}$ at large scales, to $\textrm{Fr} \approx 10^{-1}$ at vertical scales of the order of $H=1$ km \cite{Vallis2017}. Recalling that the typical oceanic Brunt-V\"ais\"al\"a frequency is $N \approx 10^{-3}$ s$^{-1}$, putting units in simulation N08 with $\textrm{Fr} = 0.11$, the values of $N$ and $H = 1$ km set $L_0 = 637$ m (i.e., the numerical domain has size $4 \times 4 \times 2$ km$^3$) and $T_0 = 8000$ s. Using these units, the simulation has r.m.s.~horizontal velocities of $\approx 0.08$ m s$^{-1}$, comparable to typical velocities in the ocean of $0.1$ m s$^{-1}$ \cite{Vallis2017}. Stokes numbers considered for the particles in table \ref{tab_parts} are also within the typical values discussed for the ocean in Sec.~\ref{sec:theory}. Of course, the separation of scales in the simulations is much smaller than in realistic oceanic flows. As a result, for the particles we can only compare their characteristic timescales with those of the smallest dynamical scales of the fluid resolved in the simulation, with the main aim of validating the bounds in Sec.~\ref{sec:bound}. Attaining realistic separations of scales for geophysical flows (together with realistic values of the Prandtl number, which is here set to 1) is impossible with current computational capabilities, a limitation of this study that should be kept in mind.

\section{Numerical results \label{sec:results}}

\begin{figure}
\includegraphics[width=0.47\textwidth]{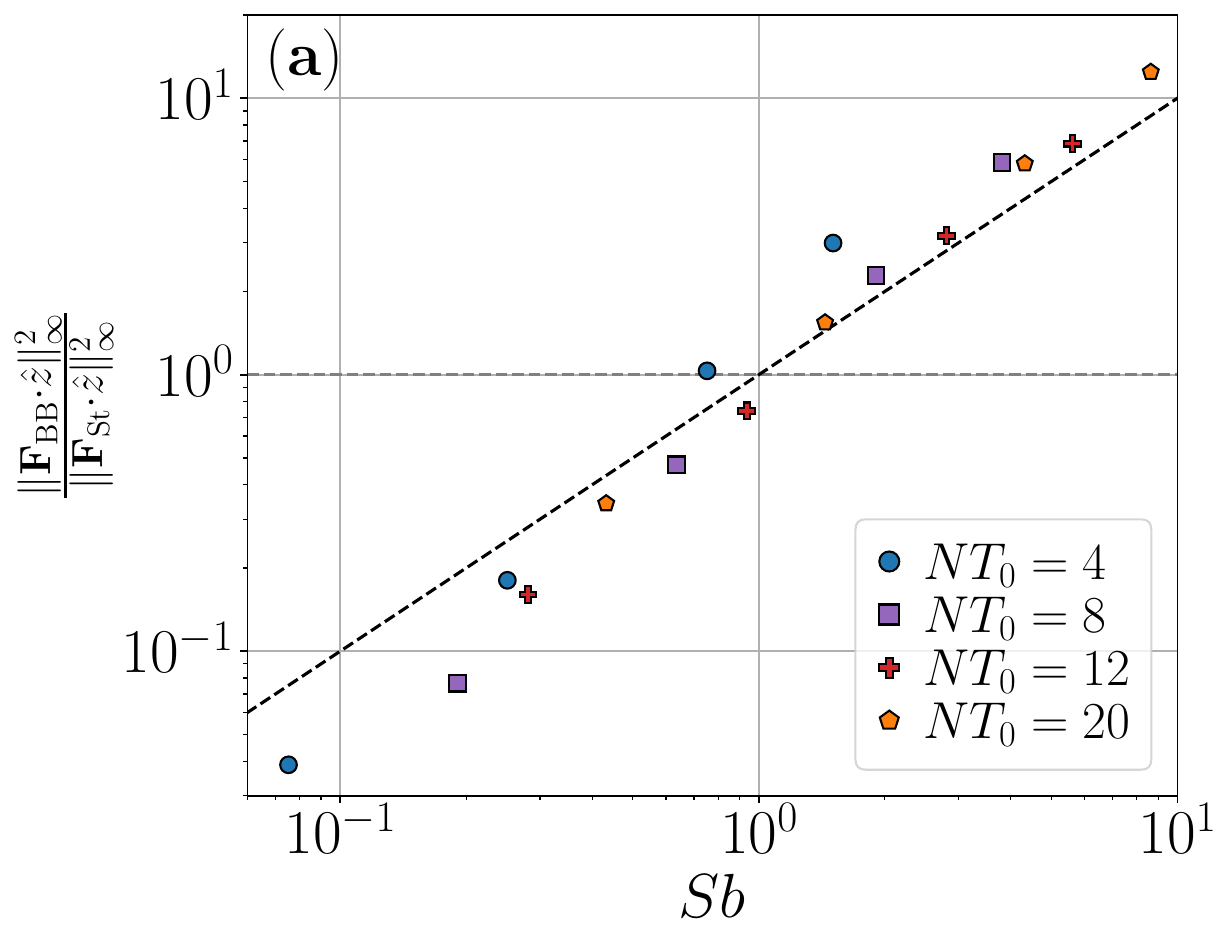}
\includegraphics[width=0.47\textwidth]{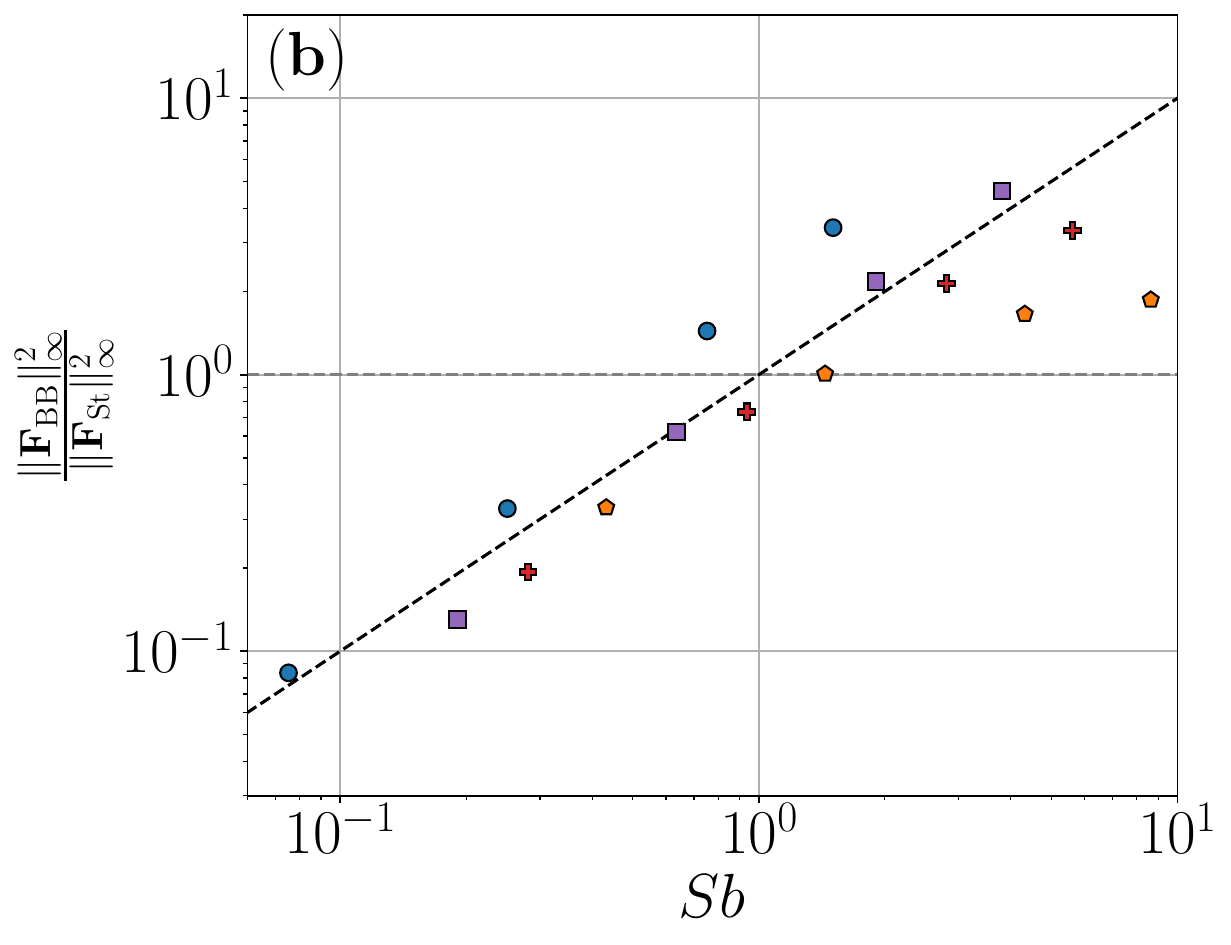}
\caption{(a) Ratio of the squared infinity norm of the vertical components of the Basset-Boussinesq force to the Stokes force, as a function of the buoyancy Stokes number $\textrm{Sb}$ for all particles and simulations. A linear relation with slope of 1 is shown as a reference. (b) Same for the infinity norms considering all components of both forces.}
\label{F_vs_sb}
\end{figure}

\subsection{Relation between the forces and the buoyancy Stokes number \label{sec:results_1}}

The first direct test of the bound given by Eq.~(\ref{cota}) considers the strength of the forces in all simulations with particles that include the Basset-Boussinesq force. To estimate $\|{\bf F}_\textrm{BB}\|_{\infty}$, $\|{\bf F}_\textrm{BB} \cdot \hat{z} \|_{\infty}$, $\|{\bf F}_\textrm{St}\|_{\infty}$, and $\|{\bf F}_\textrm{St} \cdot \hat{z} \|_{\infty}$, the maximum of the vector modulus and of the vertical component of each force were calculated for each particle in the simulations over all times, and then the average values of ${\|{\bf F}_\textrm{BB}\|^{2}_{\infty}}/{\|{\bf F}_\textrm{St}\|^{2}_{\infty}}$ and of ${\|{\bf F}_\textrm{BB} \cdot \hat{z} \|^{2}_{\infty}}/{\|{\bf F}_\textrm{St} \cdot \hat{z} \|^{2}_{\infty}}$ were computed for each value of $\textrm{Sb}$. The results are shown in Fig.~\ref{F_vs_sb}. Note that ${\|{\bf F}_\textrm{BB} \cdot \hat{z} \|^{2}_{\infty}}/{\|{\bf F}_\textrm{St} \cdot \hat{z} \|^{2}_{\infty}} < 1$ when $\textrm{St} <1$. The condition on ${\|{\bf F}_\textrm{BB}\|^{2}_{\infty}}/{\|{\bf F}_\textrm{Sb}\|^{2}_{\infty}}$ being small is also bounded by $\textrm{St}$, confirming that the vertical component of the Basset-Boussinesq force is the largest, and that it is the component of interest to obtain a bound on the effect of this force on the particles dynamics in stratified flows. Qualitatively similar results are obtained for $L^2$ norms of the forces, which are not shown (see however the next subsections for mean squared errors of various quantities with and without the Basset-Boussinesq force).

\subsection{Influence of the Basset-Boussinesq force on particle dispersion \label{sec:results_2}}

\begin{figure}
\includegraphics[width=0.46\textwidth]{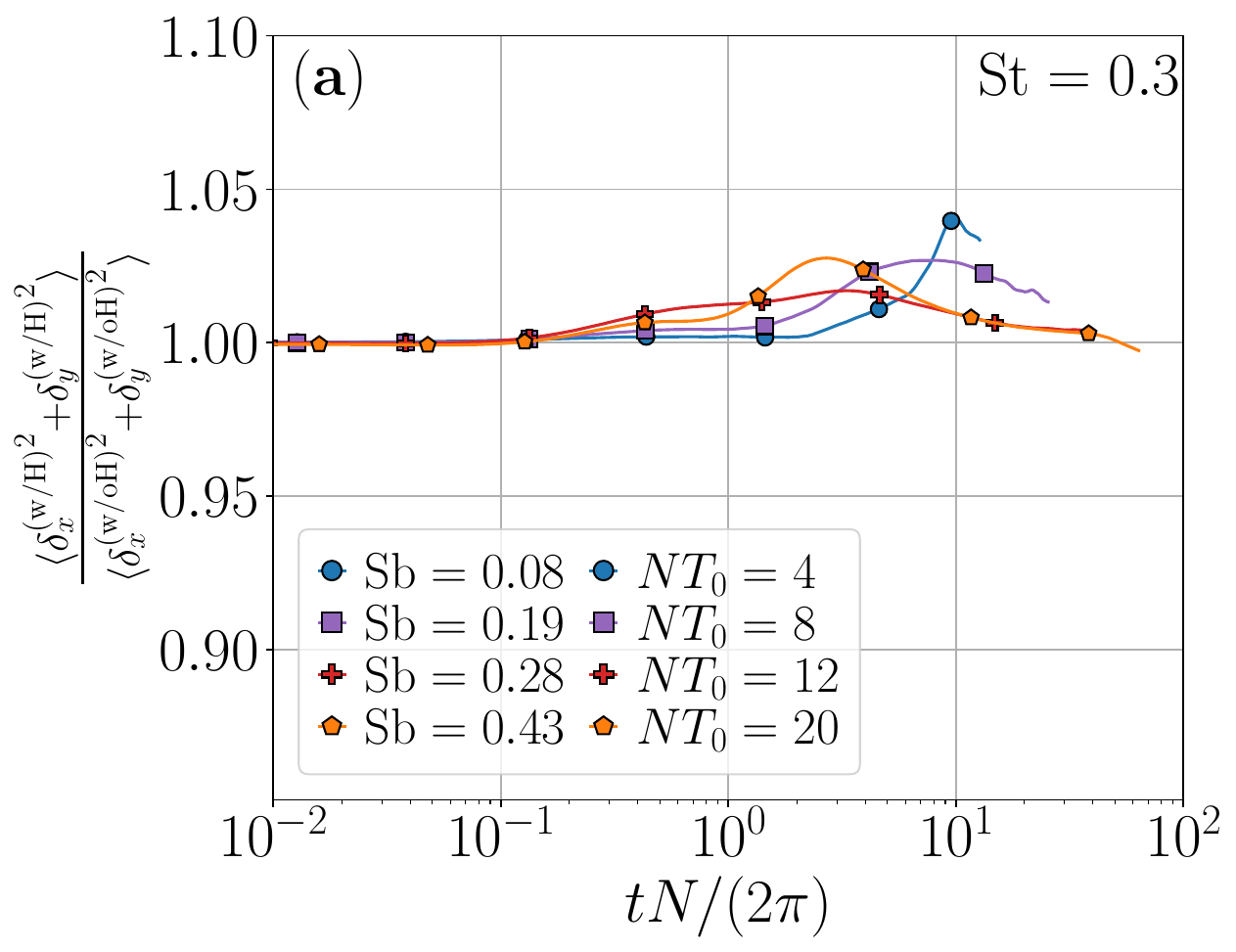}
\includegraphics[width=0.46\textwidth]{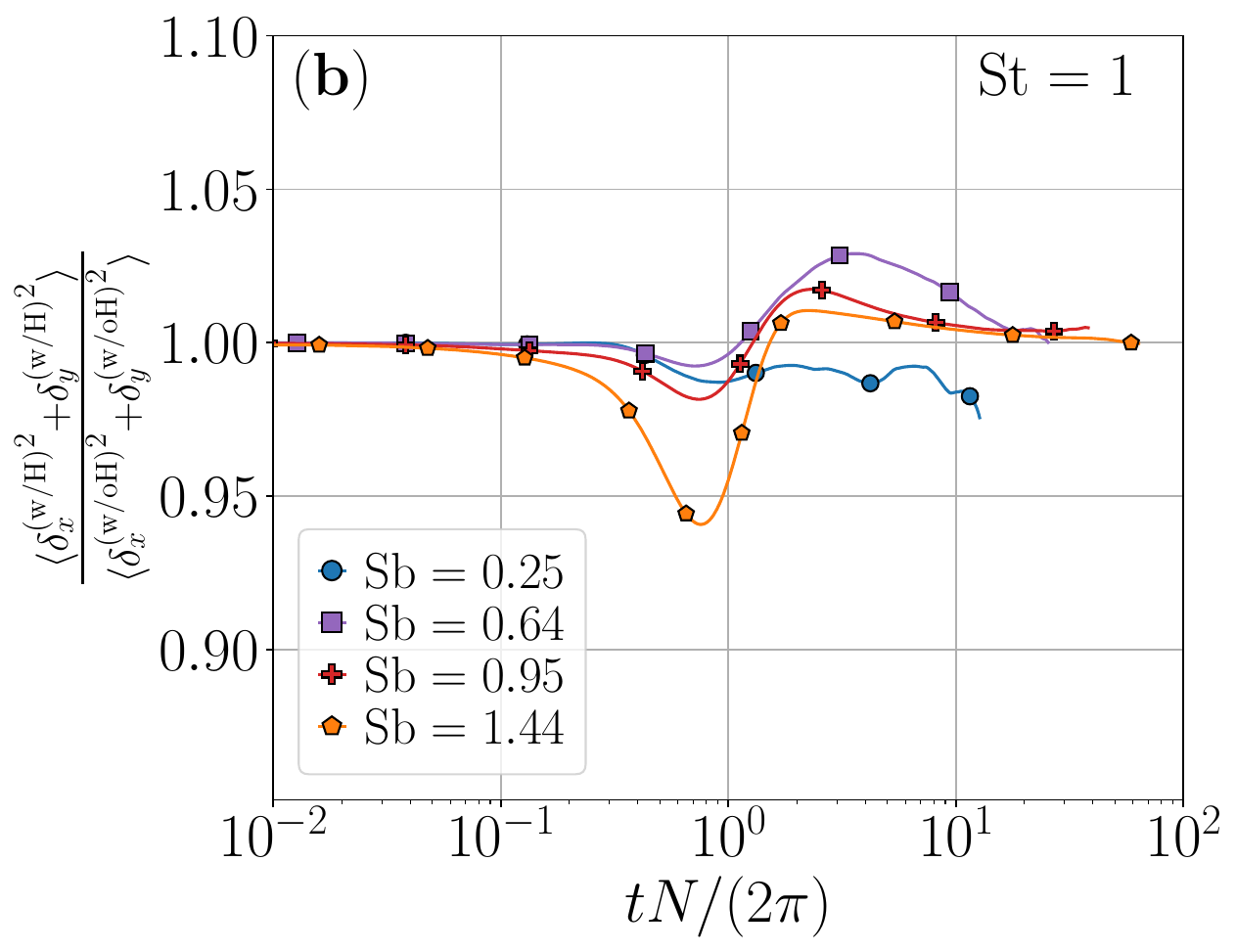}
\includegraphics[width=0.46\textwidth]{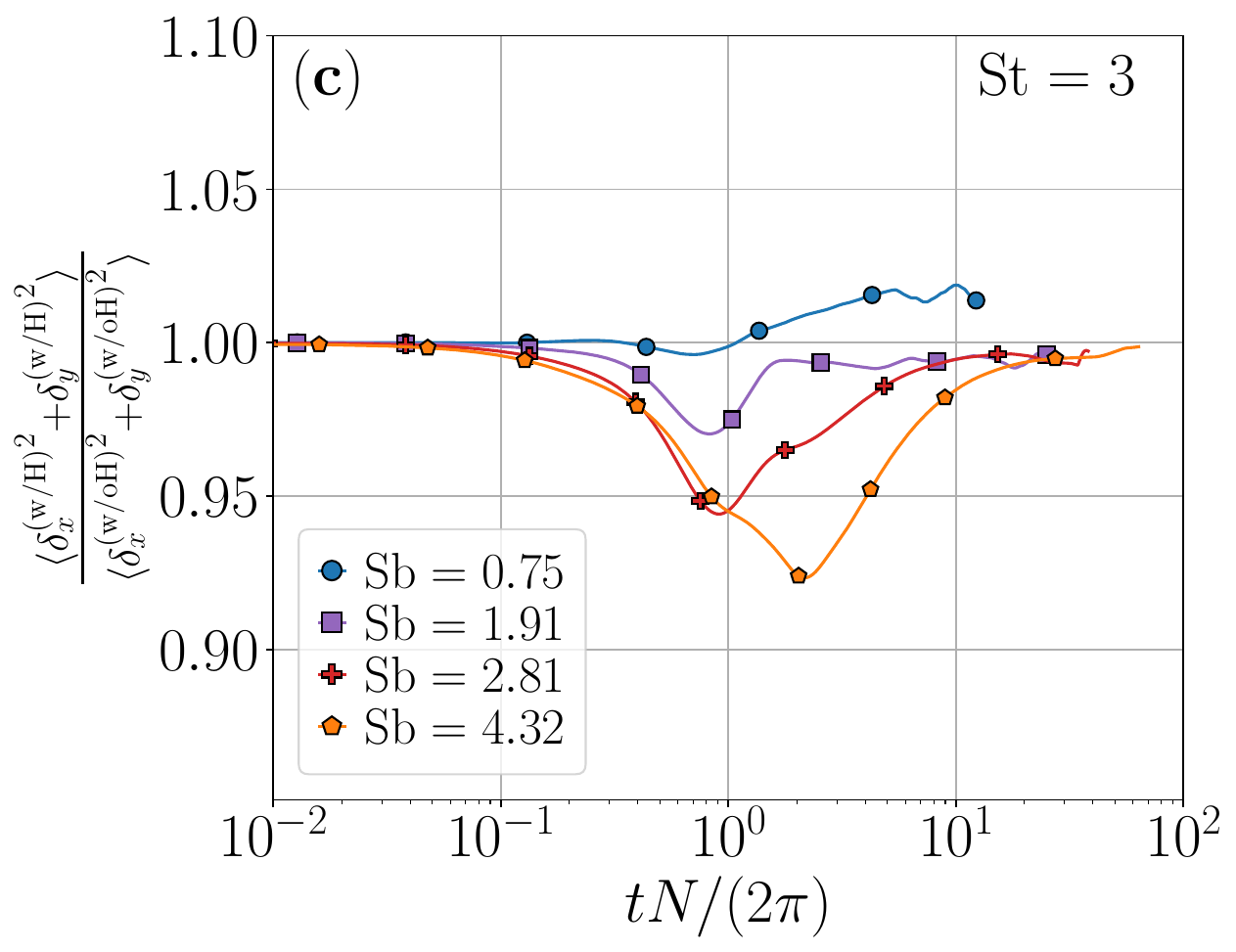}
\includegraphics[width=0.4544\textwidth]{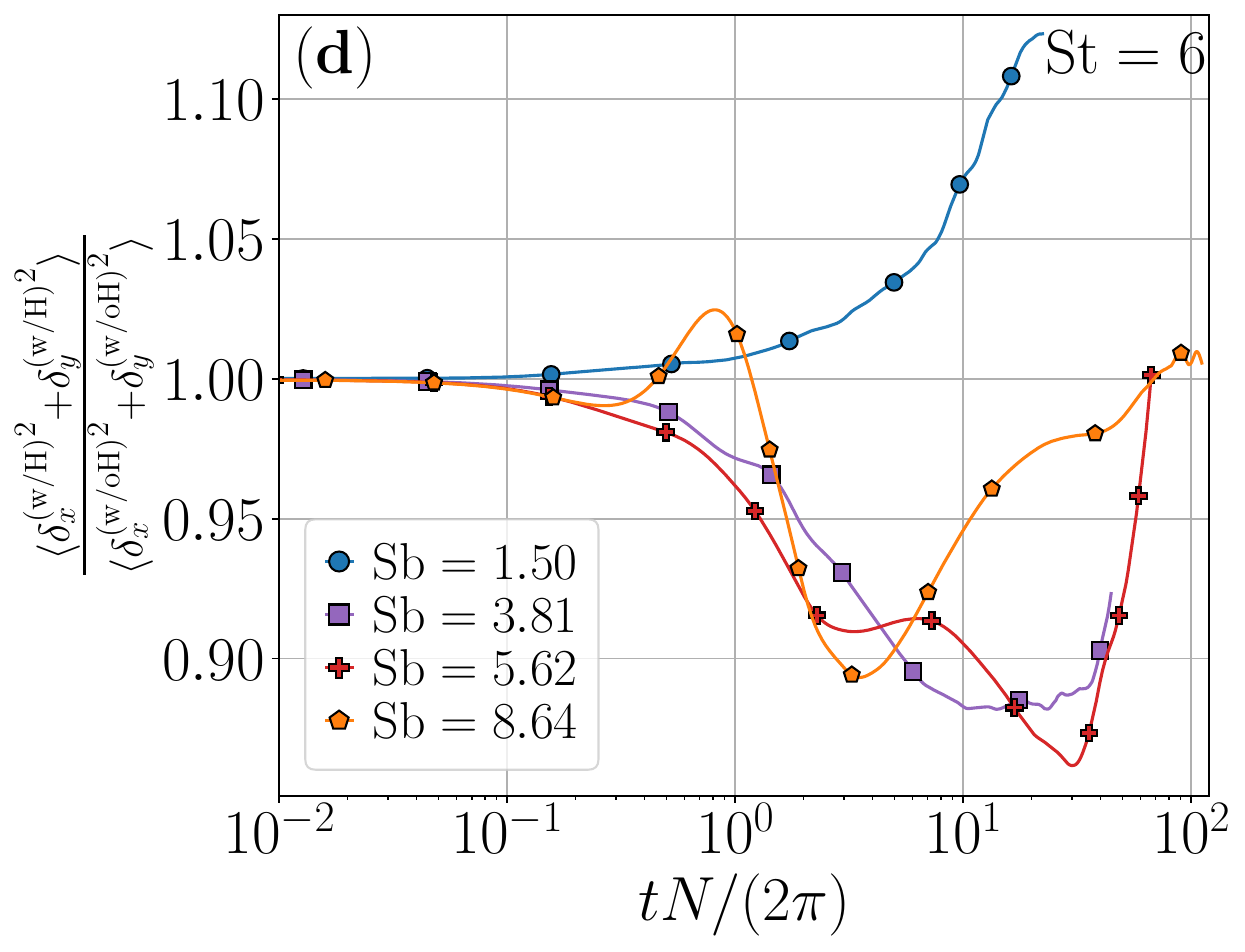}
\caption{Ratio of the mean squared horizontal dispersion considering the history term (w/H) and without the history term (w/oH), as a function of time, for particles in flows with different Froude and Stokes numbers: (a) $\textrm{St} = 0.3$, (b) $\textrm{St} = 1$, (c) $\textrm{St} = 3$, and (d) $\textrm{St} = 6$. The labels the in first panel provide the Brunt-V\"ais\"al\"a frequency for all panels. Values of $\textrm{Sb}$ for the particles are given in each panel.}

\label{disp_xlg}
\end{figure}

The next test studies the displacement of the particles in the turbulent flow, when the Basset-Boussinesq force is considered or neglected. As in the vertical direction neutrally buoyant particles are confined in a narrow layer (independently of whether the history term is present or not), we compare their mean square displacements in the horizontal direction,
\begin{equation}
    {\delta^{(j)}_i}^2(t) = \left<[x_{i}(t)-x_{i}(0)]^{2}\right>,
\end{equation}
where the subindex $i=1,2$ stands for the $x$ or $y$ coordinates, and the supraindex $j$ can be w/H or w/oH (i.e., integration with or without the history term). Figure \ref{disp_xlg} shows the ratio of the averages over all particles of the mean square horizontal displacements with and without the history term, $\langle {\delta_x^{(\textrm{w/H})}}^2 + {\delta_y^{(\textrm{w/H})}}^2 \rangle / \langle {\delta_x^{(\textrm{w/oH})}}^2 + {\delta_y^{(\textrm{w/oH})}}^2 \rangle$, as a function of time for different Froude and Stokes numbers. \ADDB{Note that in Fig.~\ref{disp_xlg}(d), it takes much longer for particles with $\textrm{St} = 6$ to reach an extremum and form a small plateau. This is due to the longer relaxation time of these particles, which results in extended transients. As previously mentioned, simulations of particles with $\textrm{St} = 6$ in this figure, as well as in the figures below, were extended to twice the duration of the other particles to ensure that the extrema were reached. However, the limitations related to the increasing difficulty in achieving a steady state as $\textrm{St}$ increases must be kept in mind as we move forward.}

For small times compared with the Brunt-V\"ais\"al\"a period, $2\pi/N$, the ratio in Fig.~\ref{disp_xlg} remains close to unity, but as time becomes close to $2\pi/N$ differences in the mean squared horizontal dispersion develop. For small values of $\textrm{St}$ (corresponding to small $\tau_p$) and large values of $\textrm{Fr}$ (i.e., small values of $N$, but still with $\textrm{Fr}<1$), the ratio hovers around unity and the differences between the particles w/H and w/oH remain below $4\%$. As the value of $\textrm{St}$ of the particles increases, the ratio deviates from unity. For fixed $\textrm{St}$ values, the ratio also increases as $\textrm{Fr}$ decreases. This increase in the differences w/H and w/oH is compatible with the increase in $\textrm{Sb}$. For large values of $\textrm{Sb}$ differences in the horizontal dispersion are between $10\%$ to $15\%$, depending on the value of $\textrm{Fr}$. However, when $\textrm{Sb}<1$, errors when neglecting the Basset-Boussinesq force consistently stay below $5\%$. It is noteworthy that in many cases the largest errors are found for intermediate times. Van Aartrijk and Clercx \cite{van_aartrijk_2010} already reported that in stratified flows different regimes develop in the vertical and horizontal dispersion of particles, with the Basset-Boussinesq force having a significant impact in the development and extension of transients at times close to the Brunt-V\"ais\"al\"a period.

\subsection{Influence of the Basset-Boussinesq force on particles velocities \label{sec:results_3}}

\begin{figure}
\includegraphics[width=0.46\textwidth]{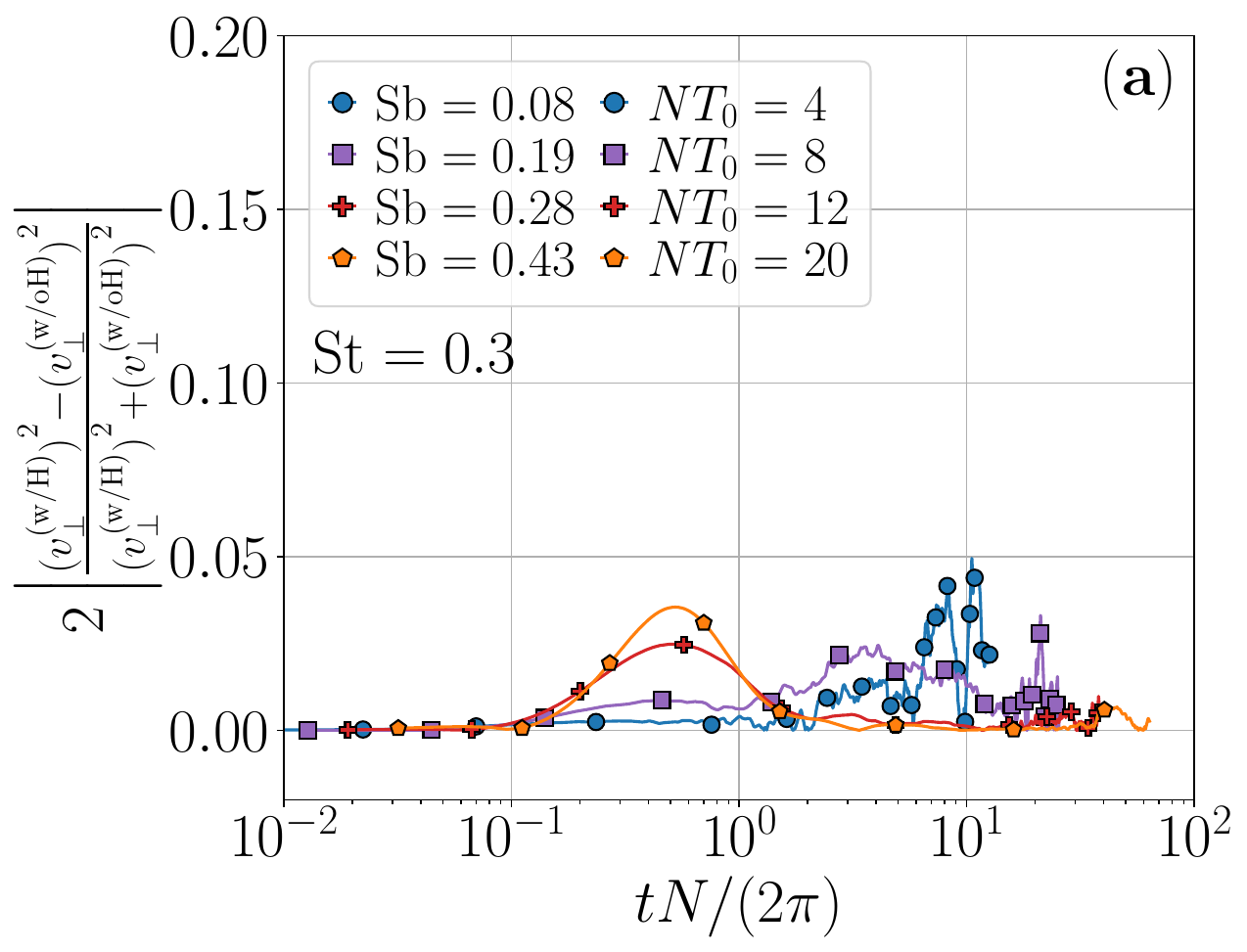}
\includegraphics[width=0.46\textwidth]{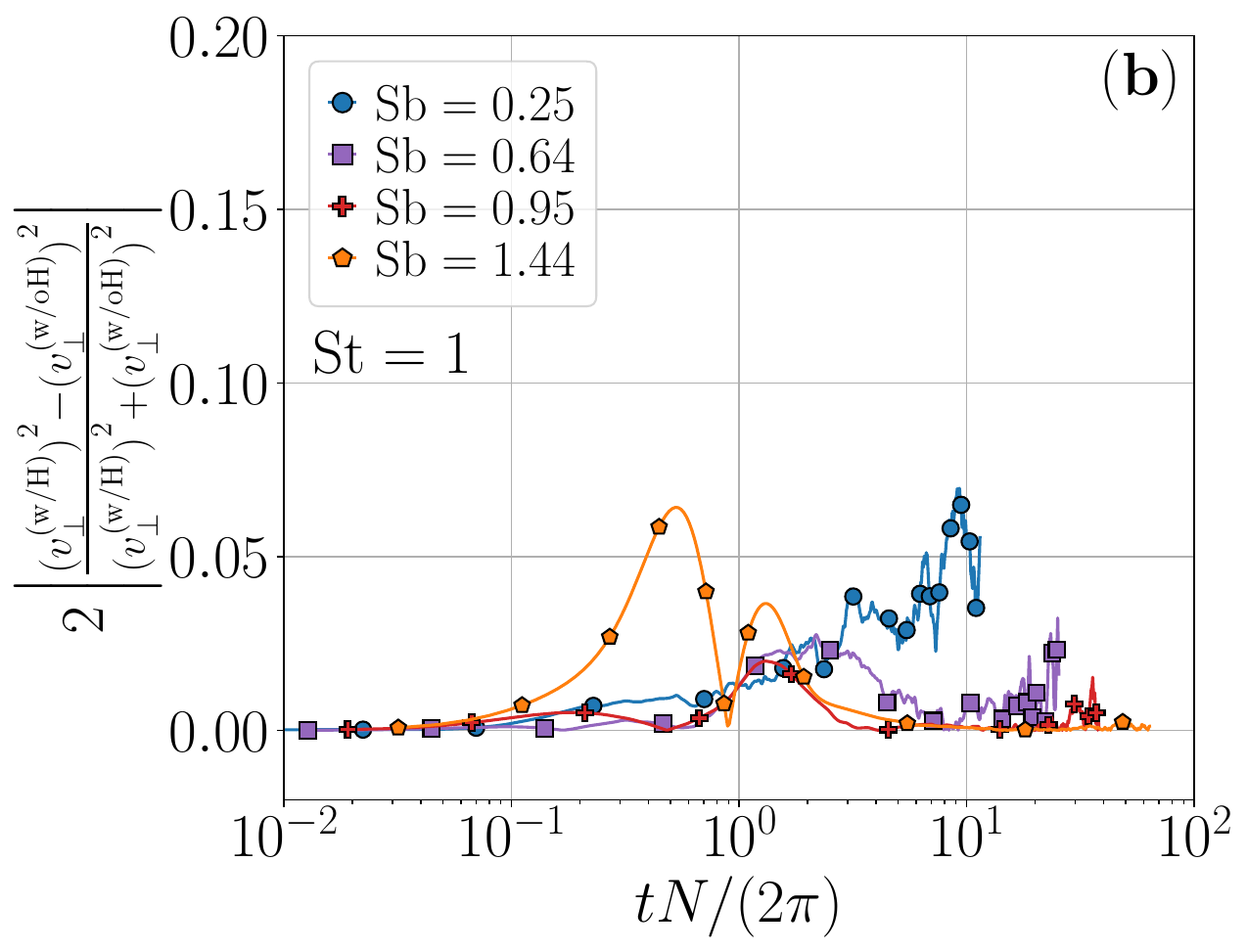}
\includegraphics[width=0.46\textwidth]{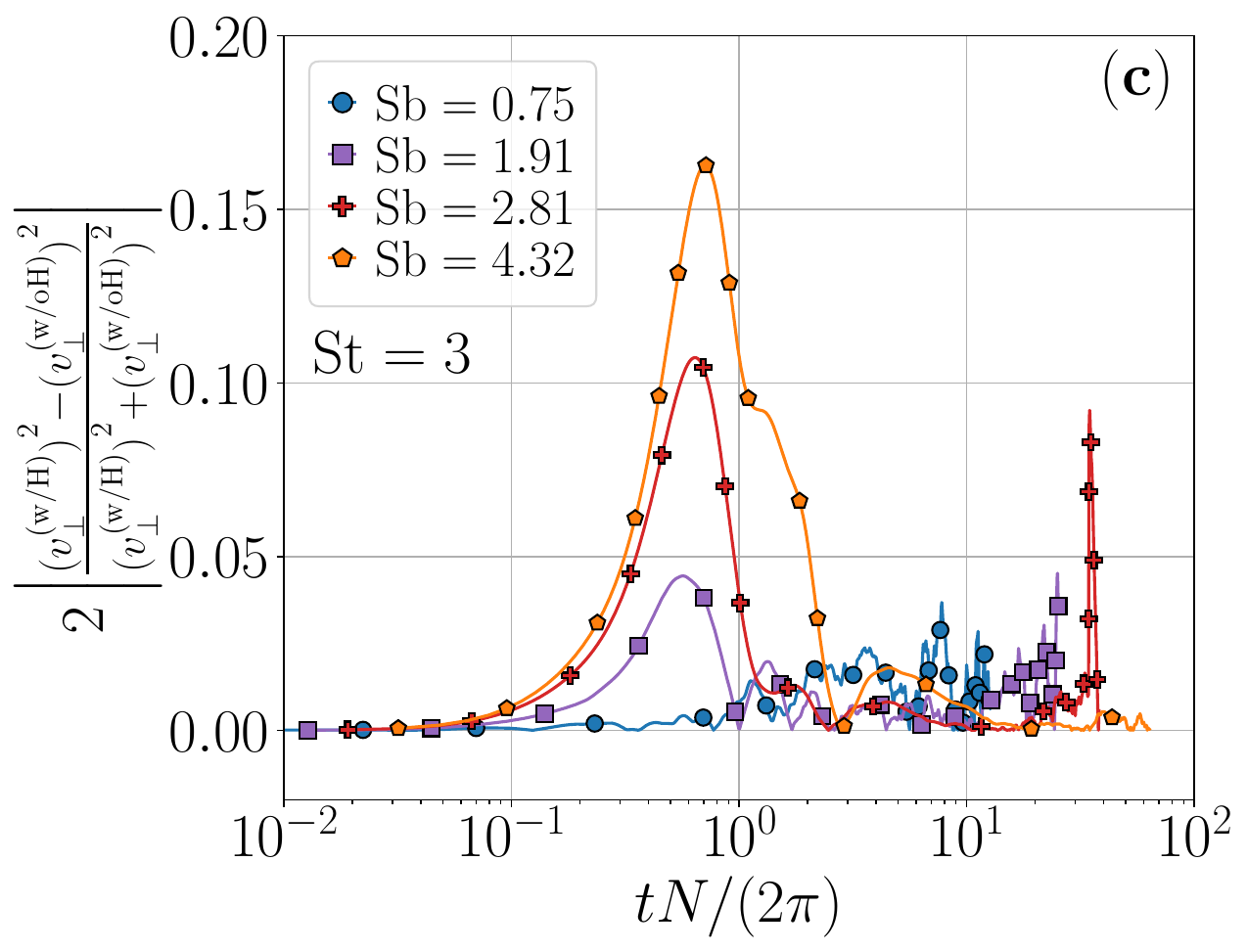}
\includegraphics[width=0.4544\textwidth]{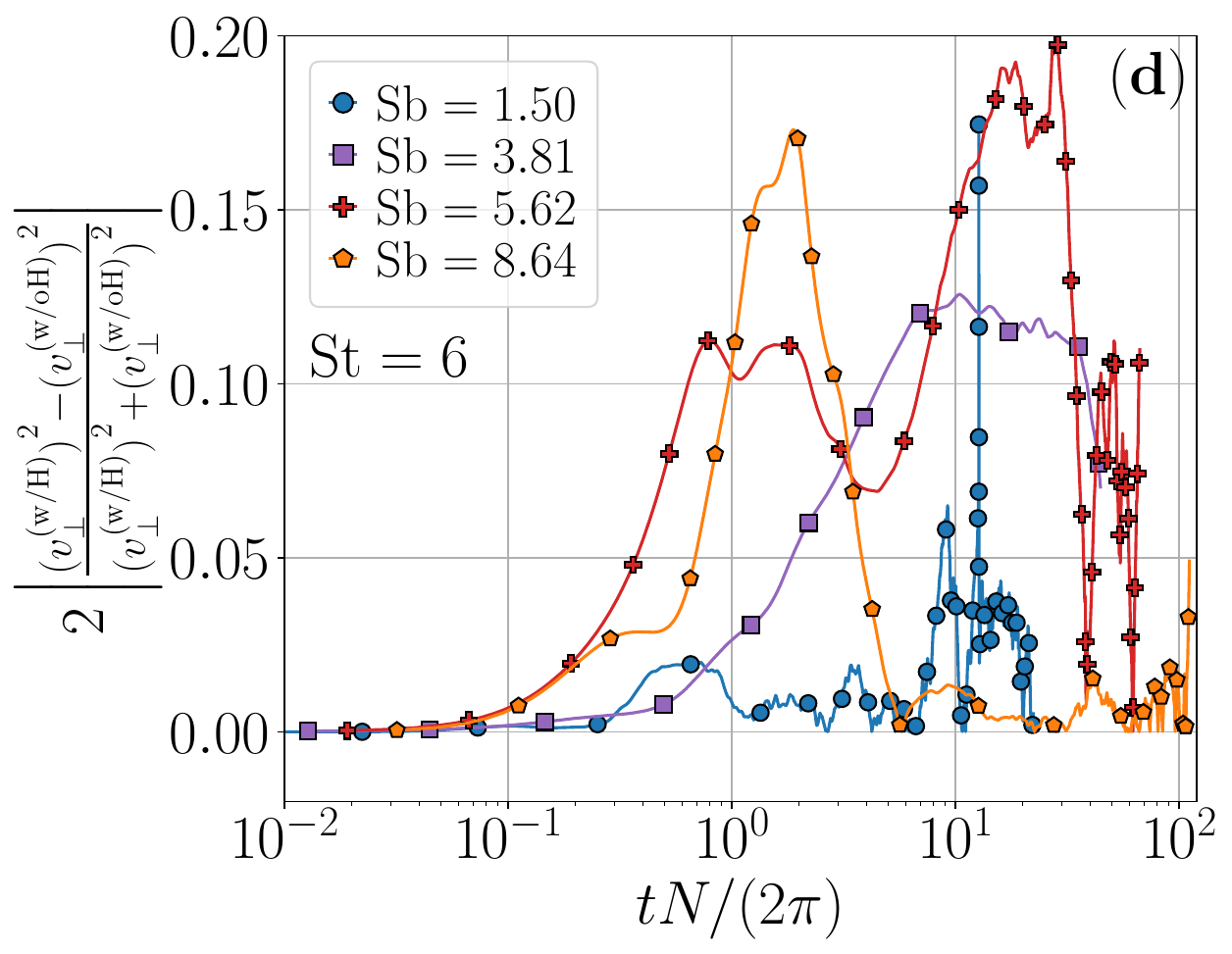}
\caption{Normalized difference between the mean squared velocity variations in cases with and without the history term (respectively w/H and w/oH), as a function of time for particles in flows with varying Froude and Stokes numbers ($\textrm{St}$): (a) $\textrm{St} = 0.3$, (b) $\textrm{St} = 1$, (c) $\textrm{St} = 3$, and (d) $\textrm{St} = 6$. The labels the in first panel provide the Brunt-V\"ais\"al\"a frequency for all panels. Values of $\textrm{Sb}$ for the particles are given in each panel.}
\label{disp_vip}
\end{figure}

We now consider the effect of the Basset-Boussinesq history term on the particles velocities. To this end we consider again the horizontal (or perpendicular) components, and we define mean squared velocity variations as
\begin{equation}
    {v_{\perp}^{(j)}}^2(t) = \left<[v_x(t)-v_x(0)]^{2}\right> + \left<[v_y(t)-v_y(0)]^{2}\right> ,
\end{equation}
where the supraindex $j$ again stands for w/H or w/oH, and where the average is computed over all particles. We can then consider the absolute value of the difference between the w/H and w/oH cases, $|(v_{\perp}^{(\textrm{w/H})})^2 - (v_{\perp}^{(\textrm{w/oH})})^2|$, normalized by the average of their mean squared values, $|(v_{\perp}^{(\textrm{w/H})})^2 + (v_{\perp}^{(\textrm{w/oH})})^2|/2$, as a function of time (see Fig.~\ref{disp_vip}).

As it was the case for the mean squared horizontal dispersion, the difference between the velocity variations with and without the Basset-Boussineq force remains small at early times, and grows as the time approaches $2\pi/N$. The error in the velocities increases with increasing Stokes number, and with decreasing Froude number. In cases with $\textrm{Sb}<1$ the error in the velocities is less than $6\%$, while in the cases in which this condition is not fulfilled the error increases up to $\approx 20\%$.
The increase in this error at intermediate times is consistent with what is observed in the horizontal particle dispersion in Fig.~\ref{disp_xlg}. Finally, note that the time of the first maximum in this error also depends on the value of $\textrm{St}$, which is consistent with observations of the time extension of the early ballistic behavior of particle dynamics in stratified flows \cite{reartes_2023}. 

\subsection{Effect of the Basset-Boussinesq force on particle clustering \label{sec:results_3}}

\begin{figure}
\includegraphics[width=0.46\textwidth]{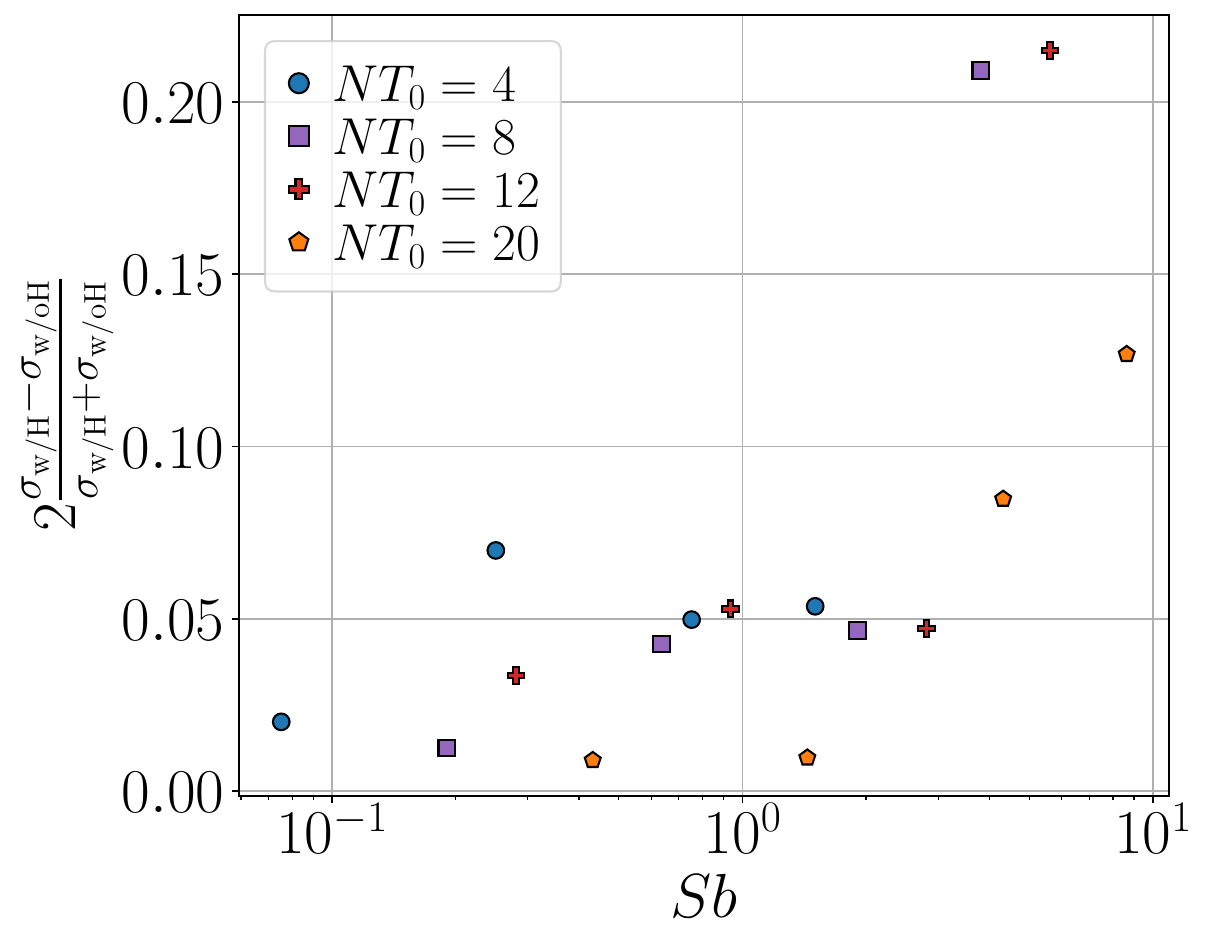}
\caption{Absolute value of the differences between the standard deviations of the Vorono\"i areas, normalized by the average of the standard deviations, for particles considering and neglecting the history term (respectively w/H and w/oH), as a function of $\textrm{Sb}$. Note the increase in the differences as $\textrm{Sb}$ increases. This trend highlights the increasing disagreement in the spatial distribution of particles and in their clustering as $\textrm{Sb}$ grows.}
\label{cluster}
\end{figure}

The vertical confinement of neutrally buoyant particles in thin layers resulting from stratification has a strong impact on inertial particle clustering \cite{reartes_2023}. This aggregation of particles is relevant in oceanic flows, in which even the dynamics of large particles are sometimes modeled using \ADDB{empirically} modified versions of the Maxey-Riley-Gatignol equation. An example we already discussed can be found in the study of Sargassum, a type of seaweed that serves as the habitat for marine fauna, but can also pose environmental challenges due to its elevated levels of arsenic and heavy metals when decomposing near coastlines \cite{BeronVera2020, AndradeCanto2022}. Another example is given by phytoplankton, that aggregates creating intricate structures spanning kilometers \cite{Martin_2003, Durham_2013}, and forming thin layers at depths correlated with regions of pronounced fluid density gradients and vertical shear, typically occurring near the bottom of the oceanic mixed layer \cite{Johnston_2009}. We thus quantify now the effect of considering or neglecting the Basset-Boussinesq force on particle clustering. To quantify particle aggregation we use a Vorono\"i tessellation. This technique has been used before to characterize the preferential concentration  of particles in laboratory experiments and in numerical simulations \cite{vor15, vor16, obligado, Obligado_2015, Sumbekova_2017, Obligado_2020, Reartes_2021, Zapata_2024}.

In a Vorono\"i tessellation each particle is assigned a cell, corresponding to all the volume (or area, in two-dimensional slices) that is closer to that particle than to any other neighbouring particle. The sizes of the cells are then inversely proportional to the particle density in that region: larger cells correspond to voids (i.e., regions with far apart particles), while smaller cells correspond to clusters (i.e, regions with particles closer to each other). The standard deviation of the volume (or area) of all the Vorono\"i cells can then we used to quantify the amount of clustering in the flow \cite{vor15, vor16}: when the standard deviation is larger than that of a random Poisson process, there is an excess of clusters and of voids compared with a random homogeneous distribution of particles. While both three- and two-dimensional tessellations have been used to study particle clustering in HIT, in the case of stably stratified flows the confinement of all particles in a layer near a the neutrally buoyant level makes it reasonable to limit the study to two-dimensional tessellations, by projecting all particles into a horizontal plane (see, e.g., \cite{reartes_2023}).

We thus computed the standard deviation of the areas of the Vorono\"i cells in simulations with the history term, $\sigma_\textrm{w/H}$, and without the history term, $\sigma_\textrm{w/oH}$. Figure \ref{cluster} shows the absolute value of the difference between these two deviations normalized by the average of the standard deviations. For $\textrm{Sb}<1$, the error when comparing cases w/H and w/oH remains below $7\%$, with most simulations having errors below $4\%$. However, for $\textrm{Sb}>1$ the error in the level of clustering grows rapidly with $\textrm{Sb}$, reaching values that exceed $20\%$. This trend highlights the increasing disagreement in particle clustering when $\textrm{Sb}>1$. Thus, the bounds obtained in Sec.~\ref{sec:bound} are also useful to estimate the conditions under which the Basset-Boussinesq force cannot be neglected when studying particle aggregation.

\section{Conclusions \label{sec:conclusions}}

We obtained a bound that is useful to determine under what conditions the Basset-Boussinesq force in the Maxey–Riley-Gatignol equation for inertial particles can be neglected, when the particles are submerged in a stratified flow \ADDB{and are at their neutrally buoyant level}. The bound is only valid for sufficiently stratified fluids with $\textrm{Fr} < 1$, i.e., it is not valid and should not be applied in the limit $N \to 0$ of isotropic and homogeneous flows. The bound motivated the definition of a buoyancy Stokes number, $\textrm{Sb} = N \tau_p$, which is the Stokes number of the particles at the fluid buoyancy scale, i.e., it is the ratio of the particle response time to the Brunt-V\"ais\"al\"a period. For sufficiently small $\textrm{Sb}$ the effect of the Basset-Boussinesq force becomes negligible.

This condition was validated using direct numerical simulations of small particles in stably stratified turbulent flows, exploring parameter space by varying the fluid Froude number and the particles Stokes number. Numerical integration of the particles was also performed considering and neglecting the Basset-Boussinesq force in the Maxey–Riley-Gatignol equation, to quantify differences between these two cases.

Using the numerical simulations we computed the infinity norm of the forces acting on the particles, and showed that the infinity norm of the Basset-Boussinesq force becomes smaller than the infinity norm of the Stokes force when $\textrm{Sb}<1$. Three other observables were considered: single particle dispersion (i.e., the mean squared distance traveled by the particles), the mean squared velocity of the particles, and particle clustering. In all cases, differences between simulations with and without the Basset-Boussinesq force were small when $\textrm{Sb}<1$, and grew rapidly with $\textrm{Sb}$ when $\textrm{Sb}>1$. Thus, the Basset-Boussinesq force must be considered in stratified flows only for particles with very large inertia, or when the stratification is too strong. The results allow the estimation of the conditions under which this force becomes relevant in different geophysical contexts \cite{Martin_2003, van_aartrijk_2008, van_aartrijk_2010, BeronVera2020, AndradeCanto2022, Ichihara_2023, OBRIEN_2023}. It is interesting that for typical parameters of oceanic flows, even particles of 50 mm radius have $\textrm{Sb} \approx 0.1$, and thus neglecting the Basset-Boussinesq force may be justified in most cases.

\bibliography{ms}

\end{document}